\documentclass[12pt]{article}
\usepackage{graphicx}
\usepackage{epsfig}
\newcommand{\beq}{\begin{equation}}
\newcommand{\eeq}{\end{equation}}
\newcommand{\beqa}{\begin{eqnarray}}
\newcommand{\eeqa}{\end{eqnarray}}
\newcommand{\beqar}{\begin{eqnarray*}}
\newcommand{\eeqar}{\end{eqnarray*}}

\newcommand{\labell}[1]{\label{#1}} %{\qquad_{#1}\label{#1}}
\newcommand{\reef}[1]{(\ref{#1})}

\newcommand{\ssc}{\scriptscriptstyle}
\newcommand{\eg}{{\it e.g.,}\ }
\newcommand{\ie}{{\it i.e.,}\ }

\newcommand{\norm}[1]{\raise.3ex\hbox{:}#1\raise.3ex\hbox{:}}

\newcommand{\al}{\alpha}

\newcommand{\veps}{\varepsilon}

\newcommand{\be}{\beta}

\newcommand{\om}{\omega}

\newcommand\tC{{\widetilde C}}

\newcommand\tF{{\widetilde F}}

\newcommand\tz{\ensuremath{\tilde{z}}}
\newcommand\tk{\ensuremath{\tilde{k}}}

\newcommand{\hk}{\widehat{k}}
\newcommand{\hH}{\widehat{H}}
\newcommand{\hG}{\widehat{G}}

\newcommand\half{\ensuremath{\frac{1}{2}}}
\newcommand\ba{\left({\beta\over\alpha}\right)}

\begin{document}

\setlength{\unitlength}{1mm}

\thispagestyle{empty}
\rightline{\small hep-th/0204144 \hfill UTPT-02-05}

%\rightline{\small McGill/02-xx \hfill }

\vspace*{2cm}

\begin{center}
{\bf \LARGE Supergravity S-Branes}\\
\vspace*{1cm}

Martin Kruczenski,\footnote{E-mail: martink@physics.utoronto.ca}
Robert C.~Myers\footnote{E-mail: rcm@hep.physics.mcgill.ca} %;\\
and Amanda W.~Peet\footnote{E-mail: peet@physics.utoronto.ca}

\vspace*{0.2cm}

{\it $^{1,3}$Department of Physics, University of Toronto}\\ {\it 60
St.~George Street, Toronto, Ontario M5S 1A7, Canada}\\[.5em]

{\it $^{1,2}$Perimeter Institute for Theoretical Physics}\\ {\it 35
King Street North, Waterloo, Ontario N2J 2W9, Canada}\\[.5em]

{\it $^2$Department of Physics, University of Waterloo}\\
{\it Waterloo, Ontario N2L 3G1, Canada}\\[.5em]

{\it $^2$Department of Physics, McGill University}\\
{\it Montr\' eal, Qu\' ebec H3A 2T8, Canada}\\[.5em]

\vspace{2cm} ABSTRACT
\end{center}
We construct supergravity solutions corresponding to space-like branes
in string theory. Our approach is to apply the usual solution
generating techniques to an appropriate time-dependent solution of the
eleven dimensional vacuum Einstein equations. In this way all
SD$p$-brane solutions are obtained, as well as the NS and M-theory
space branes. Bound states of SD$p$/SD$(p-2)$- and
SD$p$/SD$(p-4)$-branes are also constructed.  Finally, we begin an
investigation of the near-brane regions and singularity structure of
these solutions.

\vfill \setcounter{page}{0} \setcounter{footnote}{0}
\newpage

%--------------------------------------------------------------------+
\section{Introduction}

Recently space-like (and null) branes and their role in superstring
theory was discussed by Gutperle and Strominger \cite{space}.  These
investigations were motivated by the suggestion that such space-branes
may lead to a holographic duality which reconstructs a time-like
direction, just as investigations of (time-like) D-branes lead to a
space-like holography in the form of the AdS/CFT correspondence
\cite{revue}.  Such a time-like holography has been argued to exist in
the context of the de Sitter space \cite{dscft,hullfirst} and may play a role in
understanding cosmological backgrounds with the framework of string
theory.

Our understanding of the ordinary (time-like) D-branes has developed
in recent years with a wide variety of complementary descriptions.
Within perturbative string theory, D-branes can be defined as surfaces
where open strings end.  However, they also have alternative
descriptions as supergravity solutions or as solitons in the tachyonic
field theory on brane-antibrane systems. In \cite{space}, applications
of all these descriptions were discussed for space-like branes. In
particular, it was suggested that from the open string point of view,
they correspond to imposing Dirichlet boundary conditions in the time
direction.  In brane-antibrane systems they appear as time-dependent
tachyon configurations.  In supergravity they correspond to solutions
describing an incoming spherical wave packet which at later times
expands again as a spherical wave.

While much work remains to be done on the perturbative description
of space-branes, the understanding of these objects from the point
of view of low energy supergravity will be central to determining
their holographic properties. Unfortunately, only two supergravity
solutions corresponding to S-branes were presented in
\cite{space}. 
The first corresponded to an S0-brane solution of
the four-dimensional Einstein-Maxwell theory. The second solution
corresponds to the SM5-brane solution of eleven-dimensional
supergravity. The main focus of this paper is to expand the
repertoire of supergravity S-brane solutions.  In particular, we
will show that the solution-generating techniques that have been
developed within string theory can be equally well applied to this
problem. In fact, we will be able to find solutions corresponding
to all SD$p$-branes by applying various duality transformations to
a family of solutions of Einstein's equations in eleven
dimensions.  We note that similar time-dependent
solutions also appear in \cite{fern}, but these represent a special
case of the S-branes constructed here. 
The latter also seem to be closely related to
earlier cosmological solutions considered in \cite{bhstuff}.
We should add that our construction provides solutions of the standard
type II supergravities, and so these are distinct from the type II$^*$ 
E-branes considered in \cite{hullfirst,ramhull}.

The remainder of the paper is organized as follows: In section
\ref{construction}, we describe our approach of using standard duality
transformations to construct supergravity solutions corresponding to
SD$p$-branes with $p\le 6$, as well as Neveu-Schwarz and M-theory
space-branes.  In section \ref{far_and_near_limit}, we investigate the
far and near brane behaviour of the solutions. In particular, we find
certain solutions for which the scalar curvature is finite in the near
brane region. Section \ref{discussion} presents a discussion of our
new solutions. In appendix A, we construct multiply charged solutions
corresponding to SD$p$/SD($p$--2)- and SD$p$/SD($p$--4)-branes.  We
also find an embedding for the S0-brane solution of \cite{space} in
type IIa string theory as an SD6/SD0-brane.  Finally in appendix B, we
construct a type IIb supergravity solution corresponding to an
SD(--1)-brane. Throughout the paper, we use the notation of
\cite{space} where an S$p$-brane has a spatial worldvolume of $p$+1
dimensions.

%--------------------------------------------------------------------+
\section{Construction}
\label{construction}

Our strategy will be to apply the usual solution generating techniques
which arise in string theory \cite{generate}. With these tools, the
problem of solving for new solutions is reduced to an algebraic one,
rather than having to resort to solving the nonlinear differential
equations provided by Einstein's equations. We will start with an
eleven-dimensional solution of the vacuum Einstein equations
possessing the appropriate symmetries and then perform a {\it
rotation} mixing the eleventh dimension with one of space-like
dimensions. Then dimensionally reducing on the eleventh dimension
produces an SD0-brane solution of type IIa supergravity, smeared in
some number of transverse directions.  Then applying T-duality
transformations on the latter directions gives the desired SD$p$-brane
solutions. Further lifts and S-duality transformations also allow us
to construct S-brane versions of the M2- and M5-branes in M-theory and
those for the fundamental string and NS5-branes.

\begin{figure}
\begin{center}
\epsfysize=8truecm
\epsffile{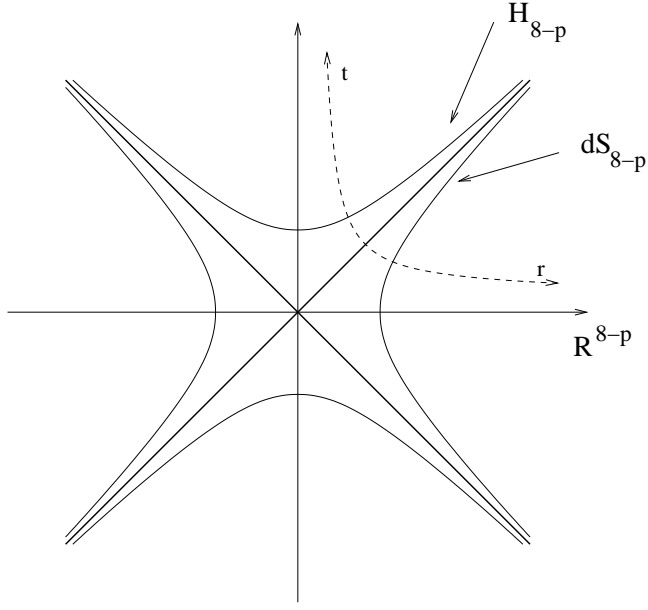}
\end{center}
\caption{Flat space parametrization appropriate for the construction
described in the text.}
\label{fig:fig1}
\end{figure}

However, in order to gain some intuition for the subsequent solutions,
we begin by presenting the following metrics on eleven-dimensional
flat space:
\beqa
ds^2&=& \left(-dt^2+t^2 dH^2_{\ssc 8-p}\right)
+\sum_{i=1}^{p+2} (d{x}^i)^2\ ,
\labell{flat1}\\
ds^2&=& \left(dr^2+r^2 d\Sigma^2_{\ssc 8-p}\right)
+\sum_{i=1}^{p+2} (d{x}^i)^2\ .
\labell{flat2}
\eeqa
Above, $dH^2_{\ssc 8-p}$ and $d\Sigma^2_{\ssc 8-p}$ represent the line
element on the (8--$p$)-dimensional hyperbolic and de Sitter spaces
with unit curvature, respectively --- we assume here and in the
following that $p\le6$.  Note that in both of these flat-space
metrics, translation and rotation invariance is manifest in the $p$+1
$x^i$ directions.  As well, there is SO(1,8--$p$) Lorentz symmetry on
either $dH^2_{\ssc 8-p}$ or $d\Sigma^2_{\ssc 8-p}$. Both of these are
symmetries which might be desired for an S$p$-brane \cite{space}.  If
we consider the (9--$p$)-dimensional subspace at, say, $x^i=0$, we see
that it is actually Minkowski space with unusual coordinates. In the
first case \reef{flat1}, the inside of the light-cone\footnote{There
are actually two patches with identical metrics corresponding to the
past and future light-cones.}  (which appears at $t=0$) is foliated by
hyperbolic surfaces with curvature $-1/t^2$. Flat space with these
coordinates is sometimes referred to as the Milne universe.  In the
second metric \reef{flat2}, the region outside of the light-cone
(which appears at $r=0$) is foliated by de Sitter space slices with
curvature $1/r^2$. This structure is illustrated in
Fig.~\ref{fig:fig1}. Our S-brane solutions will have similar
coordinates but, of course, the metric components will have extra
dependences on $t$ or $r$. However, these new solutions will approach
the above flat-space metrics asymptotically as $t,r\rightarrow\infty$.

\begin{figure}
\begin{center}
\epsfysize=8truecm
\epsffile{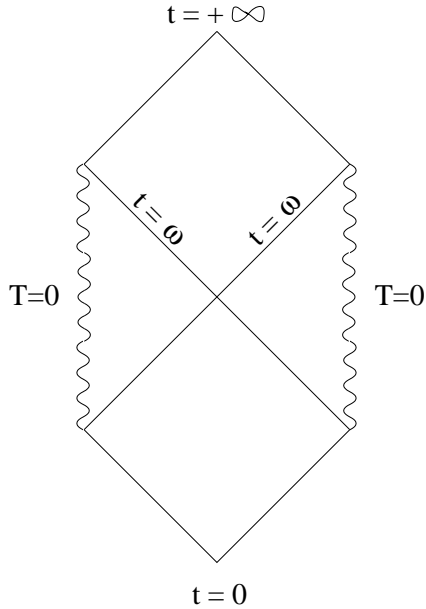}
\end{center}
\caption{The Penrose diagram for the solutions \reef{boso} with the
special choice of exponents given in eq.~\reef{bhcase}. 
Two asymptotically flat regions (at $t\rightarrow\infty$
and $t\rightarrow0$) are joined by a throat. The horizons
(at $t=\om$, $z=\pm\infty$) in the throat region surround
time-like singularities (at $T\rightarrow0$).}
\label{newfig}
\end{figure}

To begin the construction described above, we must find an appropriate
solution of the eleven-dimensional Einstein equations:
\beq
ds^2=\beta^G\alpha^H \left(-dt^2+t^2 dH^2_{\ssc 8-p}\right)
+ \left({\beta\over\alpha}\right)^{\tilde{k}} dz^2 +
\sum_{i=1}^{p+1}\left({\beta\over\alpha}\right)^{k_i}(dx^i)^2
\labell{boso}
\eeq
where
\beq
\al=1+\left({\om\over t}\right)^{7-p}\ ,\qquad
\be=1-\left({\om\over t}\right)^{7-p}\ .
\labell{boso2}
\eeq
We will assume $\om>0$ throughout the following. Further we have
singled out the coordinate $z$ anticipating that we will do a
dimensional reduction to ten dimensions on this direction.  Solving
$R_{AB}=0$ constrains the exponents to satisfy
\beqa
\tk^2+\sum_{i=1}^{p+1}k_i^2+{7-p\over4}(H-G)^2-4 {8-p\over 7-p}&=&0\ ,
\nonumber\\
\tk+\sum_{i=1}^{p+1}k_i-{7-p\over2}(H-G)&=&0\ ,
\labell{solve}\\
H+G-{4\over 7-p}&=&0\ .
\nonumber
\eeqa
This solution was found by the liberal application of `Wick rotations'
to the general solutions presented in \cite{old}. Note that most of these
solutions are singular at $t=\om$. One exception, however, is
the special case
\beq
\tk=2,\ \  H=4/(7-p),\ \  G=k_i=0,
\label{bhcase}
\eeq
for which $t=\om$ is a horizon, and the corresponding Penrose diagram
takes the form shown in Fig.~\ref{newfig}. See section 4 for further 
discussion. 

Similarly, there is a family of `exterior' solutions
\beq
ds^2=\beta^G\alpha^H \left(dr^2+r^2 d\Sigma^2_{\ssc 8-p}\right)
+\left({\beta\over\alpha}\right)^{\tilde{k}}  dz^2
+\sum_{i=1}^{p+1}
 \left({\beta\over\alpha}\right)^{k_i}(dx^i)^2\labell{bosoalt}
\eeq
where
\beq
\al=1+\left({\om\over r}\right)^{7-p}\qquad
\be=1-\left({\om\over r}\right)^{7-p}
\labell{bosoalt2}
\eeq
and the exponents are constrained as in eq.~\reef{solve}. For the special
case \reef{bhcase} (and a specific periodicity for $z$ --- see section 4),
these solutions become higher dimensional generalizations of Witten's
Kaluza-Klein bubble \cite{bubble}. In the following, we will focus on the
time-dependent solutions \reef{boso}, however, all of these constructions
follow through in the same way beginning with these solutions
\reef{bosoalt}.

%- - - - - - - - - - - - - - - - - - - - - - - - - - - - - - - - - - +
\subsection{The IIa SD0-brane}

Our basic construction was outlined in the discussion above. We
explicitly illustrate the calculations here in the construction of the
SD0-brane solution of type IIa supergravity. We begin with the above
solution \reef{boso} with $p=0$. The strategy is to then to apply a
rotation on the two coordinates, $z$ and $x^1$:
\beq
\tilde{x}^1=\cos\theta\,x^1-\sin\theta\,z
\qquad
\tilde{z}=\cos\theta\,z+\sin\theta\,x^1
\labell{rotate}
\eeq
with some fixed angle $\theta$.  Next follow this with the standard
Kaluza-Klein compactification on the $\tilde{z}$ direction, from
eleven-dimensional supergravity to ten-dimensional type IIa
supergravity:
\beq
ds^2=e^{-2\Phi/3}G_{\mu\nu}dx^\mu
dx^\nu+e^{4\Phi/3}(d\tilde{z}+C^{(1)}_\mu dx^\mu)^2
\labell{reduce}
\eeq
where $G_{\mu\nu}$ is the ten-dimensional string-frame metric.  We
note for the present case that
\beq
\left({\beta\over\alpha}\right)^{\tk}dz^2+
\left({\beta\over\alpha}\right)^{k_1}(d{x}^1)^2
={1\over F}
\left({\beta\over\alpha}\right)^{k_1+\tk}(d\tilde{x}^1)^2+F
\left(d\tz+\sin\theta\cos\theta\,{C\over F}\,d\tilde{x}^1\right)^2\ .
\labell{step3}
\eeq
where we have introduced
\beqa
C(t) &=& \left({\beta\over\alpha}\right)^{k_1}
-\left({\beta\over\alpha}\right)^{\tk}\ ,
\nonumber\\
F(t)&=&\cos^2\theta\left({\beta\over\alpha}\right)^{\tk}
+\sin^2\theta\left({\beta\over\alpha}\right)^{k_1} \ .
\labell{useful}
\eeqa

The final type IIa solution corresponding to an SD0-brane is then
\beqa
ds^2&=&F^{1/2}\beta^G\alpha^H \left(-dt^2+t^2 dH^2_{\ssc 8}\right)
+F^{-1/2}\left({\beta\over\alpha}\right)^{k_1+\tk}(dx^1)^2\ ,
\nonumber\\
\exp(2\Phi)&=&F^{3/2}\ ,
\labell{solu0}\\
C^{(1)}&=&\sin\theta\cos\theta\,{C\over F}\, dx^1\ ,
\nonumber
\eeqa
where we replaced $\tilde{x}^1 \rightarrow x^1$.  Note that in
contrast to the usual D-brane constructions, there is no extremal
limit here, \ie one can not take the limit $\omega\rightarrow 0$ while
$\cos\theta,\sin\theta\rightarrow\infty$, since the latter are
trigonometric functions (as opposed to the hyperbolic functions which
appear in constructing the usual D$p$-brane solutions). In this case,
the exponents (\ref{bhcase}) give a solution which is obviously
regular at $t=\om$.

%- - - - - - - - - - - - - - - - - - - - - - - - - - - - - - - - - - +
\subsection{SD$p$-brane solutions}
\label{sec:Spbrane}

Our starting point is now eqs.~\reef{boso} and \reef{boso2} with
general $p+1$. We do the same rotation \reef{rotate} and
compactification \reef{reduce}, as above. The resulting type IIa
solution is now
\beqa
ds^2&=&F^{1/2}
\left[\beta^G\alpha^H \left(-dt^2+t^2 dH^2_{\ssc 8}\right)+
\sum_{i=2}^{p+1}\left({\beta\over\alpha}\right)^{k_i}(dx^i)^2\right]
+F^{-1/2}\left({\beta\over\alpha}\right)^{k_{1}+\tk}(dx^{1})^2\ ,
\nonumber\\
\exp(2\Phi)&=&F^{3/2}\ ,
\labell{solupa}\\
C^{(1)}&=& \sin\theta\cos\theta\,{C\over F}\,dx^{1}\ ,
\nonumber
\eeqa
where $C(t)$ and $F(t)$ are defined as in eq.~\reef{useful}, in terms
of the $\alpha$ and $\beta$ given in eq.~\reef{boso2}.  This solution
can be regarded as an SD0-brane smeared out over the $p$ directions
$x^i$ with $i=2,\ldots,p+1$.  The SD$p$-brane solution is produced by
applying the usual T-duality transformations \cite{Tdual} on the latter 
$p$ coordinates.  The final result is
\beqa
ds^2&=&F^{1/2}\beta^G\alpha^H \left(-dt^2+t^2 dH^2_{\ssc 8-p}\right)
+F^{-1/2}\left[\sum_{i=2}^{p+1}
\left({\beta\over\alpha}\right)^{-k_i}(dx^i)^2
+\left({\beta\over\alpha}\right)^{k_{1}+\tk}(dx^{1})^2\right]\ ,
\nonumber\\
\exp(2\Phi)&=&F^{3-p\over2}
\left({\beta\over\alpha}\right)^{-\sum_{i=2}^{p+1}k_i}\ ,
\labell{solupb}\\
C^{(p+1)}&=& \sin\theta\cos\theta\,{C\over F}\,
dx^{1}\wedge \cdots\wedge dx^{p+1}\ .
\nonumber
\eeqa
Note that the general metric above is not isotropic in the worldvolume
directions \ie $x^i$ with $i=1,\ldots,p+1$. However, from a
microscopic point of view, one might expect that the supergravity
solution corresponding to a SD-brane would have an isotropic
worldvolume as there is nothing to distinguish the various directions
at this level. Isotropy in the worldvolume will be restored in the
above solution if one chooses $-k_2=\cdots=-k_{p+1}=k_{1}+\tk\equiv
n$. Notice that this excludes the case (\ref{bhcase}). If we also
define $k_{1}-\tk\equiv m$, the constraints
\reef{solve} on the exponents in this isotropic case reduce to:
\beqa
&&H={2-(p-1)n\over7-p}\ ,
\qquad
G={2+(p-1)n\over7-p}\ ,
\nonumber\\
&&9(p+1)n^2+(7-p)m^2=8(8-p)\ .
\labell{homogp}
\eeqa
The corresponding metric simplifies to
\beq
ds^2=F^{1/2}(\beta\alpha)^{2\over7-p}\left({\beta\over\alpha}
\right)^{n{p-1\over 7-p}} \left(-dt^2+t^2 dH^2_{\ssc 8-p}\right)
+F^{-1/2}
\left({\beta\over\alpha}\right)^{n}\sum_{i=1}^{p+1}(dx^i)^2\ ,
\labell{isotropical}
\eeq
while the dilaton becomes
$\exp(2\Phi)=F^{3-p\over2}({\beta/\alpha})^{pn}$.  Notice that $m$
only appears implicitly\footnote{Therefore, we have a three-parameter
family of isotropic solutions, parametrized by $(\theta,\omega,n)$.} above in
$F$ through the definition given in eq.~\reef{useful}.  We can
tabulate the maximum values that $n$ and $m$ can have for every
dimension $p=0,\ldots,6$, as follows:
\beq\labell{table1}
\begin{array}{|c|c|c|} \hline
p & n_{\rm max}=\sqrt{\frac{8(8-p)}{9(p+1)}}   &
    m_{\rm max}=\sqrt{8\frac{8-p}{7-p}} \\ \hline
0 & \frac{8}{3}                    & \frac{8}{\sqrt{7}}   \\
1 & \frac{2}{3}\sqrt{7}            & 2\sqrt{\frac{7}{3}}  \\
2 & \frac{4}{3}                    & 4\sqrt{\frac{3}{5}}  \\
3 & \frac{\sqrt{10}}{3}            & \sqrt{10}            \\
4 & \frac{4}{3} \sqrt{\frac{2}{5}} & 4 \sqrt{\frac{2}{3}} \\
5 & \frac{2}{3}                    & 2\sqrt{3}            \\
6 & \frac{4}{21}\sqrt{7}           & 4                    \\ \hline
\end{array}
\eeq
%

%- - - - - - - - - - - - - - - - - - - - - - - - - - - - - - - - - - +
\subsection{Neveu-Schwarz S-branes}

Given the SD$p$-brane solutions above, we can construct solutions
corresponding to the fundamental string and the NS5-brane, by using
S-duality in the type IIb theory. Since only Neveu-Schwarz fields are
excited in these solutions, they will also be solutions for the type
IIa, heterotic, type I and bosonic string theories.

To construct the solution for a space-like fundamental string, we
begin with the SD1-brane of the type IIb theory.  This is given by
substituting $p=1$ into the general solution \reef{solupb} of the
previous section:
\beqa
ds^2&=&F^{1/2}\beta^G\alpha^H \left(-dt^2+t^2 dH^2_{\ssc 7}\right)
+F^{-1/2}\left[\left({\beta\over\alpha}\right)^{-k_2}(dx^2)^2
+\left({\beta\over\alpha}\right)^{k_1+\tk}(dx^1)^2\right]\ ,
\nonumber\\
\exp(2\Phi)&=&F\left({\beta\over\alpha}\right)^{-k_2}\ ,
\labell{solu1b}\\
C^{(2)}&=&\sin\theta\cos\theta \,{C\over F}\, dx^1\wedge dx^2\ .
\nonumber
\eeqa
Now it is straightforward to perform an S-duality transformation to
obtain the desired SF-string.  S-duality maps \cite{john}:
$\Phi\rightarrow -\Phi$, $G_{\mu\nu}\rightarrow e^{-\Phi} G_{\mu\nu}$
and $C^{(2)}\rightarrow B^{(2)}$. The result is
\beqa
ds^2&=&\beta^G\alpha^H
\left(\frac{\beta}{\alpha}\right)^{\frac{k_2}{2}}
\left(-dt^2+t^2 dH^2_{\ssc 7}\right)
+F^{-1}\left[
\left({\beta\over\alpha}\right)^{-\frac{k_2}{2}}(dx^2)^2
+\left({\beta\over\alpha}\right)^{k_1+\tk+\frac{k_2}{2}}(dx^1)^2
\right]\ ,
\nonumber\\
\exp(2\Phi)&=&F^{-1}\left({\beta\over\alpha}\right)^{k_2}\ ,
\labell{string}\\
B^{(2)}&=&\sin\theta\cos\theta \,{C\over F}\, dx^1\wedge dx^2\ .
\nonumber
\eeqa
Note that the solution becomes isotropic in the $x^1$ and $x^2$
directions if $-k_2=k_1+\tk=n$. In this case, if we again define
$k_1-\tk=m$, the constraints \reef{solve} on the exponents yield:
\beq
H=G={1\over3}\qquad 3n^2+m^2={28\over3}\ .
\labell{homog}
\eeq
Notice here the case (\ref{bhcase}) is also excluded.

The SNS5-brane follows from S-dualizing the SD5-brane solution,
which from eq.~\reef{solupb} can be written:
\beqa
ds^2 &=& F^{\half} \beta^G\alpha^H
\left(-dt^2+t^2 dH^2_{\ssc 3}\right) +
F^{-\half} \left[ \sum_{i=2}^{6} \ba^{-k_i} (dx^i)^2 +
\ba^{k_1+\tk} (dx^1)^2\right]
\nonumber\\
e^{2\Phi} &=& F^{-1} \ba^{-\bar{k}}
\labell{sd5b}\\
F^{(3)}&=&4(k_1-\tk)\sin\theta\cos\theta\,\om^2\,\veps(H_3)
\nonumber
\eeqa
where $\bar{k}=\sum_{i=2}^{6} k_i$ and $\veps(H_3)$ denotes the volume
three-form on the hyperbolic plane $H_3$ with unit curvature.  Rather
than presenting the six-form RR potential $C^{(6)}$ as in
eq.~\reef{solupb}, we have presented the field strength of the dual
two-form potential, \ie $F^{(3)}=dC^{(2)}=*dC^{(6)}$.  The S-duality
transformation then yields the SNS5-brane:
\beqa
ds^2 &=& F \ba^{\half\bar{k}} \beta^G\alpha^H
\left(-dt^2+t^2 dH^2_{\ssc 3} \right)
+ \left[ \sum_{i=2}^{6} \ba^{\half\bar{k}-k_i} (dx^i)^2
+ \ba^{k_1+\tk+\half\bar{k}} (dx^1)^2\right]
\nonumber\\
e^{2\Phi} &=& F \ba^{\bar{k}}
\labell{sns5}\\
H^{(3)}  &=& 4(k_1-\tk)\sin\theta\cos\theta\,\om^2\,\veps(H_3)
\nonumber
\eeqa
where $H^{(3)}=dB^{(2)}$ is the NS three-form field strength.  In this
case, demanding isotropy in the worldvolume directions produces the
same constraints on the exponents as for the SD5-brane.  The latter
are given by eq.~\reef{homogp} with $p=5$.

As mentioned above, since no RR fields appear in the solutions
\reef{string} and \reef{sns5}, they will be equally valid as low
energy solutions of the type IIb or IIa, heterotic, type I or even
bosonic string theories. One could also generalize the S-duality
transformation above to a general $SL(2,R)$ mapping \cite{john}, which
produce space-branes with both Neveu-Schwarz and RR fluxes.  So in
particular, eq.~\reef{string} would be generalized to a space-like
$(p,q)$-string.

%- - - - - - - - - - - - - - - - - - - - - - - - - - - - - - - - - - +
\subsection{M-theory S-branes}

In M-theory there are also space-like counterparts to the usual M2-
and M5-branes. These can be obtained by lifting the solutions for the
space-like fundamental string \reef{string} and the SD4-brane from
eq.~\reef{solupb} to eleven dimensions.

Thus the SM2-brane is given by:
\beqa
ds^2&=&F^{\frac{1}{3}}\beta^G\alpha^H \left(\frac{\beta}{\alpha}
\right)^{\frac{k_2}{6}}\left(-dt^2+t^2 dH^2_{\ssc 7}\right)
\nonumber\\
&&\qquad\qquad
+F^{-\frac{2}{3}}
\left[\left({\beta\over\alpha}\right)^{-\frac{5}{6}
k_2}(dx^2)^2
+\left({\beta\over\alpha}\right)^{k_1+\tk+\frac{k_2}{6}}(dx^1)^2
+\left(\frac{\beta}{\alpha}\right)^{\frac{2k_2}{3}}dz^2
\right]\ ,
\labell{M2}\\
A^{(3)}&=&\sin\theta\cos\theta\,{C\over F}\,
dx^1\wedge dx^2\wedge dz\ ,
\nonumber
\eeqa
where $z$ denotes the eleventh dimension. Here $C$ and $F$ are again
defined as in eq.~\reef{useful}, and
\beq
\al=1+\left({\om\over t}\right)^{6}\ ,\qquad
\be=1-\left({\om\over t}\right)^{6}\ .
\labell{defluck}
\eeq
The exponents are constrained as in eq.~\reef{solve} with $p=1$.
Hence the unique choice for an isotropic brane is $G=H=1/3$,
$k_1=-\tk=\pm\sqrt{7/3}$ and $k_2=0$, in which case the metric becomes
\beq
ds^2=F^{\frac{1}{3}}(\beta\alpha)^{\frac{1}{3}}
\left(-dt^2+t^2 dH^2_{\ssc 7}\right)
+F^{-\frac{2}{3}} \left[(dx^1)^2+(dx^2)^2+dz^2\right]\ .
\labell{isom2}
\eeq
The case (\ref{bhcase}) gives rise to an anisotropic brane but it is interesting
to note that, in this case, dimensional reduction on $x^1$ gives rise to an
isotropic IIa Sf-string in ten dimensions. While the ten-dimensional metric
is singular but it seems that the lift to eleven dimensions may evade the
singularity at $t=\om$ \cite{liftsing} --- we consider this point in more detail
in section 4.

Similarly the SM5-brane follows from lifting the SD4-brane solution to
eleven dimensions:
\beqa
ds^2 &=& F^{\frac{2}{3}} \beta^G\alpha^H \left(\frac{\beta}{\alpha}
\right)^{\frac{\bar{k}}{3}}\left(-dt^2+t^2 dH^2_{\ssc 4}\right)
\nonumber\\
&&\qquad\qquad
+ F^{-\frac{1}{3}}\left[ \sum_{i=2}^5 \left(\frac{\beta}{\alpha}
\right)^{\frac{\bar{k}}{3}-k_i} (dx^i)^2 +
\left(\frac{\beta}{\alpha}\right)^{k_1+\tk+\frac{\bar{k}}{3}} (dx^1)^2
+\left(\frac{\beta}{\alpha}\right)^{-\frac{2}{3}\bar{k}}dz^2\right]\ ,
\labell{M5}\\
F^{(4)} &=& -6\sin\theta\cos\theta\,(k_1-\tk)\,\om^3\ \veps(H_4)
\nonumber
\eeqa
where we defined $\bar{k}=\sum_{i=2}^5 k_i$. Also $\veps(H_4)$ denotes
the volume form of the four-dimensional hyperbolic plane, \ie
$dH^2_{\ssc 4}$.  Choosing $k_{2,3,4,5}=0$, $k_1=-\tk=\pm\sqrt{8/3}$,
$G=H=2/3$ produces an isotropic brane. In this case, the metric
simplifies to
\beq
ds^2 =  F^{\frac{2}{3}}  \left(\alpha\beta\right)^{\frac{2}{3}}
          \left(-dt^2+t^2 dH^2_{\ssc 4}\right)
       +F^{-\frac{1}{3}} dx_{(6)}^2
\labell{M5h}
\eeq
where $dx_{(6)}^2$ denotes the metric for flat six-dimensional
Euclidean space.  This isotropic solution should be the same as the
SM5-brane found in \cite{space}. In fact it can be shown that the
above solution takes the form given in \cite{space} after the change
of coordinates $t^{-3}=\tanh(3\tilde{t}/2)$. We also observe that
as for the SM2-brane, considering the case (\ref{bhcase}) and dimensionally
reducing in $x^1$ gives rise to an isotropic SD4-brane.

%--------------------------------------------------------------------+
\section{Far and near brane limits}
\label{far_and_near_limit}

%- - - - - - - - - - - - - - - - - - - - - - - - - - - - - - - - - - +

It is interesting to study the structure of the SD$p$-brane
solutions\footnote{The results are very similar for the SF1-, SNS5-,
SM2- and SM5-branes.}  in the asymptotic ($t\rightarrow\infty$) and
near-brane ($t\rightarrow\om$) regions. We will address each of these
in turn. In doing so, however, we will only consider the isotropic
case where the exponents satisfy the constraints given in
eq.~\reef{homogp}. Hence in the following, recall that for these
isotropic S-branes, we have $n=k_1+\tk$ and $m=k_1-\tk$.

\subsection{Far away}

Being far from the brane corresponds to taking
$t\rightarrow\infty$. In such a limit, the SD$p$-brane metric becomes
the flat metric of (\ref{flat1}). The corrections are easy to
compute. First we have:
\beqa \frac{\beta}{\alpha} &\simeq&
1-2\left(\frac{\omega}{t}\right)^{7-p}
\nonumber\\
F &\simeq& 1- ({n}+{m}\cos2\theta)
\left(\frac{\omega}{t}\right)^{7-p}
\labell{user}\\
C &\simeq& 2 {m} \left(\frac{\omega}{t}\right)^{7-p}
\nonumber
\eeqa
Then we can find the metric, dilaton and RR potential:
\beqa
ds^2 &\simeq& \left(1-\left[
\frac{3}{2}\frac{p+1}{7-p}{n}
+\frac{{m}}{2}\cos2\theta\right]
\left(\frac{\omega}{t}\right)^{7-p}\right)
(-dt^2 + t^2 dH_{8-p}^2) + \nonumber\\
&& + \left(1-\left[
\frac{3}{2}{n}-\frac{{m}}{2}\cos2\theta\right]
\left(\frac{\omega}{t}\right)^{7-p}\right)dx_{(p+1)}^2
\labell{farry}\\
\Phi &\simeq& -\frac{1}{4}
\left[3(p+1){n}+{m}\cos2\theta\right]
\left(\frac{\omega}{t}\right)^{7-p} \nonumber\\
C^{(p+1)} &=& 2{m}
\sin\!\theta\cos\!\theta
\left(\frac{\omega}{t}\right)^{7-p}
dx^1\wedge\cdots \wedge dx^{p+1}
\nonumber
\eeqa
where in the metric, $dx_{(p+1)}^2$ indicates the line element on flat
($p$+1)-dimensional Euclidean space.  Note that the power of $t$ in
the RR potential is such that the surfaces of constant $t$ will carry
a constant flux, that is, $*F^{(p+2)}\propto \veps(H_{8-p})$. In fact,
the latter result applies for the full nonlinear solution, as was
explicitly shown for the SD5-brane in eq.~\reef{sd5b}.  The above form
of the asymptotic solution also indicates that we are using the
Feynman propagator, that is, the analytic continuation of the
Euclidean propagator $1/r^{7-p}$. We will return to this point in the
discussion section.

%- - - - - - - - - - - - - - - - - - - - - - - - - - - - - - - - - - +
\subsection{Close up metric}

Naively, the flat space metric \reef{flat1} would indicate that we
approach the brane, \ie, the tip of the light-cone, in the limit
$t\rightarrow0$ (with fixed coordinates on the hyperbolic space).
However, it is clear from the solutions \reef{solupb} that this
intuition must be modified in the case of the strongly gravitating
branes. We see that components of the metric may vanish or diverge at
$t=\omega$. Hence the most interesting limit when we get close to the
brane is $t\rightarrow \omega$ where $\beta\rightarrow0$.  In fact
given the latter, it is useful to use $\beta$ as a coordinate instead
of $t$. Using this new coordinate, the metric for the isotropic
(SO($p+1$)-invariant) SD$p$-branes becomes
\beqa
ds^2 & = & \sqrt{F(\beta)}\beta^G(2-\beta)^H
{\frac{\omega^2}{(1-\beta)^{2/(7-p)}}} \left[
-{\frac{ d\beta^2}{(7-p)^2(1-\beta)^2}} + dH_{8-p}^2 \right]
\nonumber\\
& & \qquad+ {\frac{1}{\sqrt{F(\beta)}}}
\left({\frac{\beta}{2-\beta}}\right)^{n} dx_{(p+1)}^2 \ .
\eeqa
For calculating the curvature at small $\beta$, it suffices to take
\beq
ds^2\simeq {2^H\omega^2} \sqrt{F(\beta)}
\beta^G \left[-{\frac{
d\beta^2}{(7-p)^2}} + dH_{8-p}^2 \right] +
{\frac{1}{\sqrt{F(\beta)}}}
\left({\frac{\beta}{2}}\right)^n dx_{(p+1)}^2
\eeq
where
\beq\label{effmn}
F(\beta)=\cos^2\!\theta
\left({\frac{\beta}{2}}\right)^{\half({n}-{m})}
 +  \sin^2\!\theta
\left({\frac{\beta}{2}}\right)^{\half({n}+{m})}\ .
\eeq

\begin{figure}
\epsfysize=6truecm\epsfbox{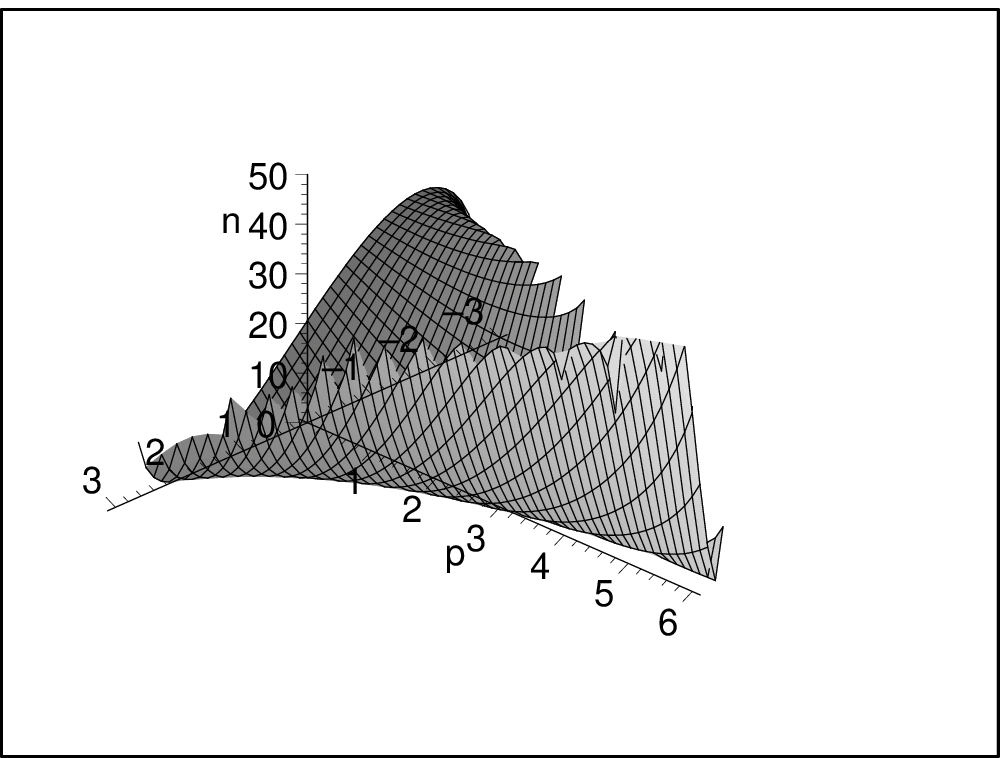}
\epsfysize=6truecm\epsfbox{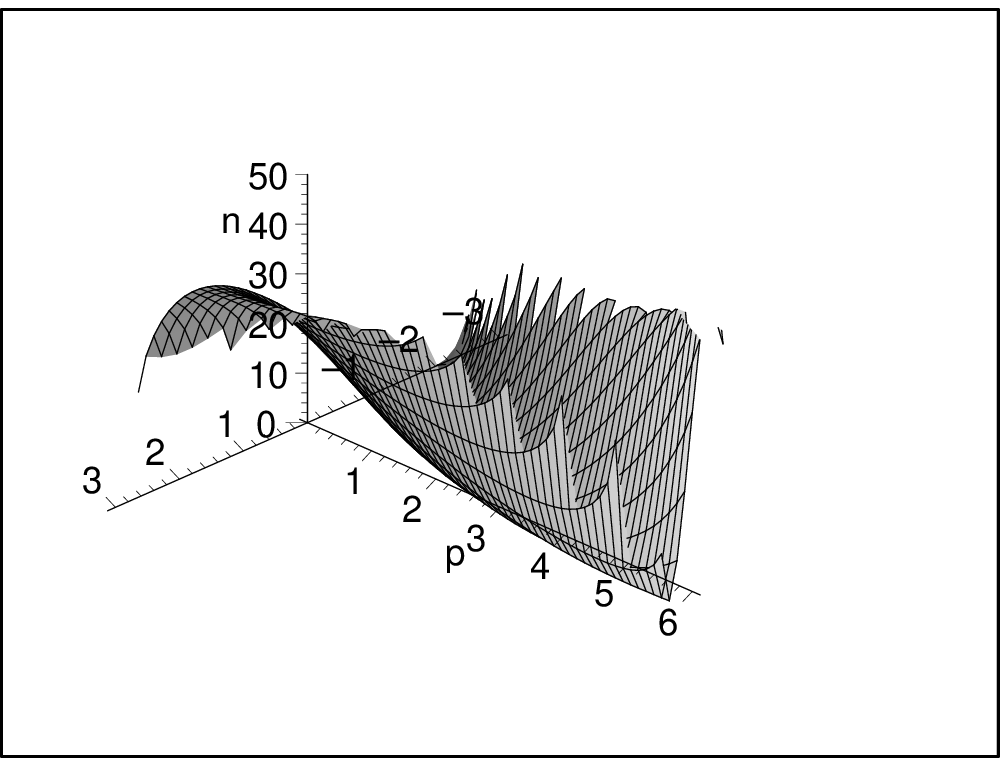}
\caption{\small The function ${\cal{R}}_p(n)$, as a function of
${n},p$ for ${m}>0$ and ${m}<0$ respectively. (The edges of the
surfaces occur where $n$ runs out due to the reality constraint on
$m$.)}
\label{fig:figs23}
\end{figure}

At small $\beta$, we find the Ricci scalar to be
\beq
R \simeq {\frac{(7-p)^2}{2^{2+H}\omega^2}}\
{\frac{{\cal{R}}_p(n)}{\beta^{2+G} \sqrt{F(\beta)}}} \eeq
where eq.~\reef{homogp} requires that $(7-p)G=2+(p-1)n$ and
$(7-p)H=2-(p-1)n$.  Then, with the abbreviation ${\dot{\ }}\equiv
\beta (d/d\beta)$, the function ${\cal{R}}_p(n)$ can be written as
\beqa
{\cal{R}}_p(n) & = &
\left(\frac{\ddot{F}}{F}\right)\left[2(7-2p)\right] +
\left(\frac{\dot{F}}{F}\right)^2 \left[(p-3)(p+1)\right] +
\left(\frac{\dot{F}}{F}\right)\left[n(-4p^2+12p-2)\right]
\nonumber\\
&& + {\frac{1}{(7-p)}} \left[-4(8-p)+n^2(-4p^3+28p^2+4p+8)\right]
\eeqa
Eq.~\reef{homogp} also fixes $m$ with $(7-p){m}^2 = 8(8-p)-9(p+1)n^2$.
Let us assume for definiteness that ${m}>0$, \ie take the
positive\footnote{For ${m}<0$, we simply swap
$\cos\!\theta\leftrightarrow\sin\!\theta$ and
$+{m}\leftrightarrow-{m}$ in the following.}  root for $m$ in the
constraint equation.  Then by eq.~(\ref{effmn}), the first term in $F$
and its derivatives will dominate at small $\beta$.  Next, we need to
inspect ${\cal{R}}_p(n)$ in the expression for $R$.  All derivatives
of $F$ appearing here are of the same order as $F$ itself.  In fact,
at leading order in $\beta$, ${\dot{F}}/F=(n-m)/2$, and
${\ddot{F}}/F=(n-m)^2/4$.  This gives ${\cal{R}}_p(n)$ as a function
of $n$ and $p$ only.  For illustrative purposes, we plot this function
for both signs of $m$ in Fig.~\ref{fig:figs23}.

The function ${\cal{R}}_p(n)$ has a zero, so for all $p$ there is a
solution with $R{=}0$ at $\beta\rightarrow 0$.  The root of
${\cal{R}}_p(n)$ can be solved for analytically:
\beq\label{nplusminus}
n_\pm(p) = \pm {\frac{(p-3)}{6}}\sqrt{\frac{2(8-p)}{(p+1)}}
\quad = \pm{\frac{(p-3)}{4}} \left| n_{\rm max} \right|
\eeq
where $n_{\rm max}$ is the maximum allowed value of $n$, tabulated in
(\ref{table1}).  Also, as can be seen from Fig.~\ref{fig:figs23}, the
magic value $n_-$ must be chosen for $m>0$, and $n_+$ for $m<0$.

For generic values of $n,p$, however, ${\cal{R}}_p(n)$ will not
vanish.  Then, the behaviour of $R$ at small $\beta$ is governed by
\beq
R \simeq {\frac{1}{\omega^2}}
{\frac{1}{\beta^{2+G}\sqrt{F(\beta)}}}
\simeq {\frac{1}{\omega^2}} {\frac{1}{\cos\!\theta}}
\beta^{-\delta_g}
\eeq
where the exponent $\delta_g$ is given by
\beq
\delta_g =
{\frac{[8(8-p)+3n(1+p)-{m}(7-p)]}{4(7-p)}}
\eeq
By expressing $m$ in terms of ${n}$ and $p$ via the constraint
\reef{homogp}, it can be shown that $\delta_g$ is strictly positive.
This implies that $R$ blows up at $\beta\rightarrow 0$ generically.

The SNS-branes and SM-branes can be analyzed similarly.  The
expression for $R$ is almost identical, except for different powers of
$F$ and $\beta$ in the denominator.  $\delta_g$ is still strictly
positive.  We find zeroes in ${\cal{R}}$ as follows: SF1:
$n_\pm=\pm\sqrt{7}/3$, SNS5: $n_\pm=\pm 1/3$, SM2: $n_\pm=0,\pm 4/3$,
and SM5: $n_\pm=0$.  Note in particular the root $n=0$ for the SD3,
SM2, and SM5. Also note that for $p=0$ the root $n=2$ corresponds
to the case (\ref{bhcase}).

To demonstrate that a solution is nonsingular, however, it is not
sufficient to show that $R{=}0$.  It is tedious to check the
$R^{\mu\nu}R_{\mu\nu}$ and
$R^{\mu\nu\lambda\sigma}R_{\mu\nu\lambda\sigma}$ invariants.  We have
explicitly computed both, using the full metric, for the $n=0$ cases
(SD3, SM2, SM5).  We find that $R^{\mu\nu}R_{\mu\nu}$ is zero at
$\beta\rightarrow 0$, but that
$R^{\mu\nu\lambda\sigma}R_{\mu\nu\lambda\sigma}$ blows up there.
Therefore, these solutions are singular, if in a somewhat milder way.
This was most likely inevitable due to the fractional powers appearing
in the metric, even for $n=0$.

%- - - - - - - - - - - - - - - - - - - - - - - - - - - - - - - - - - +
\subsection{Close up dilaton and R-R fields}

We can also consider the close-up dilaton behaviour for the
SD$p$-branes.  We find that
\beq e^\Phi \sim \left(\cos\!\theta\right)^{(3-p)/2}
\left({\frac{\beta}{2}}\right)^{-\delta_\Phi} \eeq
where the exponent $\delta_\Phi$ is
\beq
\delta_\Phi = {\frac{1}{8}} \left[{m}(3-p) - 3n(p+1)\right]
\eeq
Interestingly, $\delta_\Phi$ is zero for the magic $n_\pm$ of
(\ref{nplusminus}), for the same choice of ${\rm{sgn}}(m)$:
\beq
\delta_\Phi = 0 \,,\quad \left\{
\begin{array}{l}
n=n_-(p) \,,\quad m>0\cr
n=n_+(p) \,,\quad m<0
\end{array}
\right.
\eeq

For the R-R field of the SD$p$-branes,
\beq
C_{1\cdots p+1} =  \frac{\cos\!\theta\sin\!\theta \
C(\beta)}{F(\beta)}
=  \frac{\cos\!\theta\sin\!\theta \left[
\beta^{m}-(2-\beta)^{m} \right]}
{\left[\cos^2\!\theta(2-\beta)^{m} +
\sin^2\!\theta\beta^{m}\right]}
\eeq
Again picking ${m}>0$ for definiteness, we find:
\beq C^{p+1} \simeq -\tan\!\theta \left[ 1 -
\sec^2\!\theta\left({\frac{\beta}{2}}\right)^{m} \right]
dx^1\wedge \ldots \wedge dx^{p+1} \eeq
%
%Near $\beta=0$, $d/d\beta \sim -(\omega/(7-p)) d/d\tau$, so for the
%special case ${m}=1$, the R-R field strength for $\beta\rightarrow 0$
%is constant(!).

%--------------------------------------------------------------------+
%\newpage
\section{Discussion}
\label{discussion}

In this paper, we have constructed a wide variety of space brane
solutions for supergravity equations using the standard solution
generating techniques available in string theory \cite{generate}.  In
particular, we have obtained maximally symmetric SD$p$-brane solutions
preserving $SO(p+1)\times SO(8-p,1)$. Note that for each value of $p$,
there is a three parameter family of these isotropic solutions. From a
microscopic point of view, it is not clear why there should be three
parameters rather than just one corresponding to the number of
branes. Note that in contrast the isotropic S-branes arising in
M-theory have two parameters ($\theta$, $\omega$). This reduction for 
the isotropic SM-branes could in principle be used to determine a preferred set of
solutions for the SD$p$-branes.

Central to our construction was finding an appropriate family of
generating solutions given in eq.~\reef{boso}, which correspond to
anisotropic solutions of Einstein's equations in eleven dimensions.
Nearly all of these solutions contain curvature singularities at
$t=\om$. As a result, most of the resulting S-brane solutions contain
at least mild curvature singularities, as discussed in section 3.
However, there remains work to be done in understanding the near-brane
regions of the full set of S-brane solutions which we have
constructed. In any event, the pervasive appearance of singularities,
as well as the extra parameter in the isotropic solutions, suggest
that not all of these solutions will be relevant for string theory.
That is, reasonable physical sources will not be found in string
theory for all of the solutions presented in this paper.

A better behaved set of generating solutions arises with the choice
of exponents given in eq.~\reef{bhcase}. In this case, $t=\om$ becomes
a horizon, as is easily seen by making the coordinate transformation:
$T=\alpha^{2/(7-p)}t$. In this case, the metric reduces to
\beq
ds^2=-f^{-1}(T)\,dT^2+T^2dH^2_{8-p}
+f(T)\,dz^2+\sum_{i=1}^{p+1}(dx^i)^2
\labell{nicer}
\eeq
where $f(T)=1-4\left({\om/ T}\right)^{7-p}$
and the Penrose diagram takes the form given in Fig.~\ref{newfig}.
Hence there are two asymptotically flat regions (at $t\rightarrow\infty$
and $t\rightarrow0$) joined by a throat. One also finds time-like singularities
(at $T\rightarrow0$) behind the horizons (at $t=\om$) in the throat region.
With these exponents \reef{bhcase} in any of the solutions of section 2, one has
an anisotropic S-brane which, however, inherits the same casual structure. This
special class of solutions corresponds to those presented in \cite{fern}.
Similar solutions were also considered in \cite{cornal}.

Of course, the above remarks apply to section 2.4 and so choosing eq.~\reef{bhcase}
produces an anisotropic SM2-brane. However, as noted there, one could make
a dimensional reduction on the $x^1$ direction to produce an isotropic IIa
SF-string in ten dimensions. Naively then while the ten-dimensional metric
is singular, it seems that the lift to eleven dimensions would evade the
singularity at $t=\om$ \cite{liftsing}. However, one must be cautious because
implicitly we are choosing $x^1$ to be periodic. Hence in the eleven-dimensional
solution, this circle shrinks to zero size at $t=\om$ to produce
a `conical' singularity. The near `horizon' geometry in the $t$ and $x^1$ directions
is essential flat space. In this limit, it may be possible to understand the
periodicity of $x^1$ as orbifolding by a particular boost symmetry
\cite{cornal}, \cite{olden}, \cite{new}, \cite{nikita}. Hence M-theory
may be able to resolve this singularity, however, one must note that
the region behind the `horizon' seems as though it must be problematic
since it contains closed time-like curves.

Further progress in identifying the physically relevant
supergravity solutions would come from a better understanding of
the microphysical, \ie the perturbative string, picture of the SD-branes.
The suggestion of \cite{space} is that S-branes would appear in
perturbative string theory by introducing an open string sector with
Dirichlet boundary conditions in the direction of time.\footnote{An
open string theory on null-branes was investigated by \cite{kogan}.}
Thus these strings are confined to a space-like surface at a given instant
of time. The worldvolume theory on such an S-brane seems to be
an euclidean Yang-Mills theory. The scalar corresponding to transverse
displacements in the time direction would have a negative kinetic
term. Such theory appears to contain negative norm states, unless
they can be eliminated by means of the gauge fixing. A related unusual
property of the field theory would be that the Lorentz symmetry
$SO(8-p,1)$ of the transverse space becomes the `R-symmetry' group,
which is hence non-compact.  The same symmetry is maintained in our
supergravity solutions as the isometry group of the hyperbolic space
$H_{8-p}$.

Recall that there is a class of supergravity solutions generated from
eq.~\reef{nicer} which is somewhat better behaved at $t=\om$.
However, these space branes are anisotropic with the $x^1$ direction
being distinct from the other worldvolume directions, $x^i$ with $i=2,\ldots,p+1$.
It could be that this particular direction
is singled out by imposing different boundary conditions on the
worldvolume fields. In part, we are motivated to make this suggestion
by two observations: First, the generating solution \reef{nicer} is roughly
a `Wick rotation' of a black brane solution. In the context of time-like
D-branes, such nonextremal supergravity solutions correspond to having
the worldvolume theory at finite temperature. The latter can be evaluated
by considering a path integral where antiperiodic boundary
conditions are imposed on the fermions in the euclidean time direction.
Second, as noticed in section 2 (see further discussion below), for the
analogous exterior solutions \reef{bosoalt}, the corresponding coordinate
closes off smoothly at $t=\om$, which results in antiperiodic boundary
conditions for the supergravity fermions at infinity. Again similar considerations in
the context of time-like D-branes (see, \eg \cite{hott}) would dictate
that the worldvolume fermions are also antiperiodic. In any event, if
this speculation is correct, our supergravity results would indicate
that a Scherk-Schwarz compactification of the worldvolume theory is
better behaved than the uncompactified theory.

As noted in \cite{space}, the perturbative SD-branes would naturally
source on-shell closed strings in the surrounding bulk
spacetime. However, there is an ambiguity in these long-range bulk
fields which can alternatively be seen as an ambiguity in the Green's
function or an ambiguity in the boundary conditions for the bulk
fields. In examining the asymptotic late time-behaviour in section
3.1, the fields displayed a $1/t^{7-p}$ behavior. The latter would
naturally be associated with the Feynman propagator in Minkowski
space, which is the analytic continuation of the usual Euclidean
Green's function.  Enforcing causality, on the other hand, would
require the use of the retarded and advanced Green's function. The
advanced one would describe fine-tuned incoming radiation that creates
the space-brane, and the retarded one would describe the outgoing
radiation after the brane disappears~\cite{space}.  The retarded and
advanced Green's functions are different from the Feynman propagator
and also have
different properties in even and odd dimension, \eg for even
dimensions and massless fields, they only have support on the
light-cone. From a causal perspective, the Feynman propagator would
contain extra homogeneous solutions which neither help to create the
brane nor come from its decay.  The possible role of these bulk fields
in a brane/bulk duality remains mysterious to us.  The apparent
advantage of using the Feynman propagator, however, is that solutions
of any dimensionality appear to be on more of an equal footing.

Moreover it is not clear to us if supergravity solutions corresponding
to the retarded plus advanced propagator actually exist. In fact,
in~\cite{space}, it was argued that the creation of the brane can be
described as the excitation of the open string tachyon in an unstable
brane or in a brane/anti-brane system. This tachyon field then decays
to a (possibly new) vacuum generating an outgoing pulse of
radiation. We expect that a supergravity solution can be used to
describe this process only when a large number of branes is
created. However in that case one might expect that the incoming wave
collapses to form a black hole. While it may still excite the tachyon
field, the latter stages of this process would be hidden behind an
event horizon. This picture is reminiscent of the recent discussion
presented in~\cite{Dani}. There it was argued that tachyon lumps
corresponding to a large number of branes reside inside their own
Schwarzschild radius and consequently, can not decay by classical
radiation.  The preferred process for large $N$ appears to be the
creation of open strings living on the branes (as this process is
enhanced by a factor of $N$) rather than closed string emission.  For
a more detailed discussion of this issues we refer the reader
to~\cite{Dani}.  Here, we only note that this discussion shows that,
if the physics of these supergravity solutions is contained in the
theory describing open strings with Dirichlet boundary conditions in
the time direction, then such theory should contain a rich structure
with quite different regimes at small and large $N$ as well as an
ambiguity which corresponds to the choice of boundary conditions in
supergravity.  It is clear that more work is needed in order to
properly understand the worldvolume theory and translate its
properties into supergravity statements. In the present paper we
assumed that the world volume theory can be defined in such a way as
to correspond to the Feynman propagator which on the other hand seems
natural since the latter is defined just by an analytic continuation
from the Euclidean one. Note also that the solutions presented
in~\cite{space} which contain gravity are also of the same type.

It is interesting that other time-dependent backgrounds have recently
been studied from a closed string worldsheet point of view using
orbifolds \cite{new}, \cite{nikita}, \cite{cornal}, \cite{vijay},
\cite{newer}, \cite{newerr}.  The
time-dependence is generated by involving the time direction in one of
the orbifolding operations.  These studies may provide useful insights
to understanding the spectrum of open string theory living on an
SD-brane.

Our approach to constructing SD$p$-brane solutions worked in a
straightforward way for $0\le p\le6$. One might consider extending
this family of solutions beyond this range of $p$. In appendix B, we
construct a solution which seems to correspond to an SD(--1)-brane in
type IIb supergravity. Note that this solution is {\it not} the usual
instantonic D(--1)-brane, but rather a time-dependent solution in
ten-dimensional Minkowski-signature spacetime. It seems that one can
construct solutions for $p=7$ and 8 using essentially the same
procedures as in section 2. The case $p=8$ follows straight forwardly
from the replacement $p=8$ in all expressions and the case $p=7$ can
be described somewhat loosely speaking as a limiting procedure $p =
7-\epsilon$ with $\epsilon\rightarrow 0$.  In light of the previous
discussion, the cases $p=6,7$ and 8 are particularly interesting
because they involve less than three {\it space-like} transverse
dimensions. Hence an incoming shell would not be expected to form a
black hole horizon.  To finish this discussion, we add that one can
not have an SD9-brane in ten-dimensional Minkowski space, \ie a
nine-brane would necessarily fill the time direction.

In this paper, we focused on the time-dependent solutions generated
from eq.~\reef{boso}, as these naturally seem to describe the
solutions sourced by space-branes, \ie inside the light-cone of the
S-brane source. We would also like to consider briefly the solutions
that would be generated in the identical fashion using
eq.~\reef{bosoalt} as the generating solution. These would seem to
describe solutions in a region casually disconnected from the
space-branes, \ie outside the light-cone of the S-brane source. It is
not clear what the role of these regions would be in a holographic
description of the S-branes. However, as noted above, our supergravity
solutions would seem to be related to a perturbative framework using
the Feynman propagator for the bulk fields. The latter would naturally
introduce homogeneous waves in the region outside the light-cone, and
hence perhaps these `exterior' solutions could naturally be matched on
to the S-brane solutions presented above, at least for certain choices
of the parameters. Certainly, a much more thorough investigation of
the near-brane regions would be needed before such matching of
solutions could be accomplished.

As noted before, the solutions in eq.~\reef{bosoalt} are closely
related to Witten's Kaluza-Klein bubble \cite{bubble}, which
demonstrates the instability of the KK vacuum (in certain
theories). In fact for $k_i=0$, $\tk=2$, $G=0$ and $H=4/(7-p)$, these
solutions are higher dimensional generalizations of Witten's
solution. The $z$ direction smoothly closes off at $r=\om$ if $z$ has
period $2\pi\,2^{9-p\over7-p}\om/(7-p)$. Of course, in this solution,
one could trade $z$ for for one of the $x^i$ as coordinate on the
circle that closes off. As all of the constructions of section 2
follow through unchanged if one begins with the solutions \reef{bosoalt},
and in particular the KK bubbles, one would construct `charged'
generalizations of the usual vacuum solutions. These might
provide interesting time-dependent string theory backgrounds \cite{bath}.
Note, however, as these particular
`exterior' solutions close off smoothly, they would certainly not seem
amenable to the matching suggested above.

We close with one final comment on the `exterior' solutions.  These
solutions realize the $SO(1,8-p)$ symmetry by introducing a foliation
of the transverse space in terms of de Sitter space slices. If we
examined the near-brane region, these solutions would naturally take
the form of a warped product of $dS_{8-p} \times R^{p+2}$, similar to
those discussed in \cite{hullds}.  As de Sitter space seems to arise
quite rarely in string or M-theory \cite{hull2}, these solutions
deserve further study.

%--------------------------------------------------------------------+
\section*{Acknowledgments}

We would like to thank Bobby Acharya, Chris Hull and Fernando Quevedo
for discussions. Research by AWP was supported in part by the
Cosmology and Gravity Program of the Canadian Institute for Advanced
Research (CIAR) and by the National Sciences and Engineering Research
Council (NSERC) of Canada.  MK and RCM are supported in part by NSERC
of Canada and Fonds FCAR du Qu\'ebec. RCM would like thank the Isaac
Newton Institute for Mathematical Sciences for their hospitality in
the final stages of this work. As this paper was being prepared, an
e-print \cite{gut2} appeared which constructs similar S-brane
solutions by directly solving the relevant Einstein equations.

\appendix

%--------------------------------------------------------------------+
\section{Multiply charged S-branes}

In the case of time-like D-branes, an important role is played by
bound states where branes of different dimensionality bind together to
form a single object carrying two (or more) RR charges.  A notable
example is the D1/D5 system which plays a central role in
investigations of black hole microphysics \cite{Stromi} and the
AdS/CFT correspondence \cite{revue}. Similarly one can consider
multiply charged SD-branes although it is not clear if they will play
the same essential role. In any case, to stress the simplicity of our
approach, we construct the supergravity solutions corresponding to
such multiply charged branes and leave their detailed study for future
work.

We start with SD$p$/SD$(p-2)$-brane solutions (with $p\ge 2$). Their
construction is a minor extension of the approach given in section
\ref{sec:Spbrane}. There we started from the SD0-brane smeared in $p$
transverse directions and performed $p$ T-dualities. In the present
case, we stop after two T-dualities and lift to eleven dimensions
producing a smeared SM2-brane. Then we rotate the coordinates $x^1$
and the new eleventh coordinate $z$ by a constant angle $\gamma$ and
dimensionally reduce back producing a smeared SD2/SD0-brane. Finally
we perform the last $p-2$ T-dualities. The result is
\beqa
ds^2 &=& \tF^{\half} F^{\half} \alpha^H \beta^G
(-dt^2+t^2 dH_{8-p}^2) +
\tF^{\half} F^{-\half} \left[  \ba^{-k_2} (dx^2)^2 +
\ba^{-k_3} (dx^3)^2\right] +
\nonumber\\
 &&  +  \tF^{-\half} F^{-\half} \left[\ba^{k_1+\tk} (dx^1)^2
     + \sum_{i=4}^{p+1} \ba^{-k_i} (dx^i)^2 \right]
\nonumber\\
e^{2\Phi} &=& F^{\frac{3-p}{2}} \tF^{\frac{5-p}{2}}
\ba^{-\sum_{i=2}^{p+1} k_i}
\labell{oldo}\\
B^{(2)} &=& \sin\gamma\sin\theta\cos\theta\, \frac{C}{F}
\,dx^2\wedge dx^3
\nonumber\\
C^{(p-1)} &=& \sin\gamma\cos\gamma\, \frac{\tC}{\tF}\,
              dx^1 \wedge dx^4 \wedge \ldots \wedge dx^{p+1}
\nonumber\\
C^{(p+1)} &=& \cos\gamma\sin\theta\cos\theta\, \frac{C}{F}
\,dx^1\wedge\ldots \wedge dx^{p+1}
\nonumber
\eeqa
where $F$ and $C$ denote the same functions \reef{useful}
as before and we have introduced two new
functions $\tF$ and $\tC$, defined as
\beq
\tF(t) = \cos^2\gamma + \sin^2\gamma \frac{1}{F}
\ba^{k_1+\tk+k_2+k_3}\ ,
\qquad
\tC(t) = \frac{1}{F}\ba^{k_1+\tk+k_2+k_3}-1\ .
\label{old2}
\eeq
Further note that $\alpha$ and $\beta$ are defined as in
eq.~\reef{boso2} with $p$ (and not $p$--2).  As a trivial check we can
see that when $\gamma\rightarrow 0$, we recover the
SD$p$-brane. Slightly less trivial is to check that for
$\theta\rightarrow 0$, the solution becomes a smeared
SD($p$--2)-brane. A straightforward calculation shows that this is
indeed the case after we make the replacements:
\beq
\begin{array}{l}
\hH = H - q,\ \  \hG = G+q,\ \ \widehat{\tk} = \tk-2q,
\ \ \hk_1 = k_1+q, \\
\hk_2 = -k_2-\tk+q,\ \ \hk_3 = -k_3-\tk+q, \ \
\hk_{i=4\ldots p+1} = k_i +q
\end{array}
\labell{old3}
\eeq
with $q=(\tk+k_3+k_3)/3$. The `hatted' values satisfy the conditions
(\ref{solve}) if the original ones do.
Note that these bound state solutions involve a nontrivial
Neveu-Schwarz two-form $B$, just as is found for time-like
D$p$/D($p$--2)-brane solutions \cite{bbbb} --- see, also, \cite{green}.

The SD$p$/SD$(p-4)$-brane solutions (with $p\ge4$) are constructed
in a similar manner. The final result is
\beqa
ds^2 &=& \tF^{\half} F^{\half} \left[
\beta^G\alpha^H(-dt^2+t^2 dH_{8-p})\right] +
         \tF^{\half} F^{-\half} \sum_{i=2}^{5} \ba^{-k_i} (dx^i)^2 +
\nonumber\\
     && +\tF^{-\half} F^{-\half} \left[\ba^{k_1+\tk} (dx^1)^2  +
         \sum_{i=6}^{p+1} \ba^{-k_i} (dx^i)^2 \right]
\nonumber\\
e^{2\Phi} &=& F^{\frac{3-p}{2}} \tF^{\frac{7-p}{2}}
\ba^{-\sum_{i=2}^{p+1} k_i}
\labell{bother}\\
C^{(p-3)} &=& \sin\gamma\cos\gamma\  \frac{\tC}{\tF}\
              dx^1\wedge dx^6 \wedge \ldots dx^{p+1}
\nonumber\\
C^{(p+1)} &=& \sin\theta\cos\theta\  \frac{C}{F}\
dx^1\wedge \ldots dx^{p+1}
\nonumber
\eeqa
In this case the functions $\tF$ and $\tC(t)$ are defined to be
\beq
\tF (t) = \cos^2\gamma +
\sin^2\gamma\ba^{k_1+\tk+\sum_{i=2}^{5} k_i}\ ,
\qquad
\tC (t) = \ba^{\tk+\sum_{i=1}^{5} k_i}-1\ .
\labell{bother2}
\eeq
Again the $\gamma\rightarrow0$ limit is trivial and one can easily
verify that the limit $\theta\rightarrow 0$ produces a smeared
SD$(p-4)$ solution with the replacements:
\beq
\begin{array}{l}
\hH = H - q,\ \  \hG = G+q,\ \ \widehat{\tk} = \tk-2q,
\ \ \hk_1 = k_1+q, \\
\hk_{i=2\ldots 5} = -k_i-\tk+q, \ \ \hk_{i=6\ldots p+1} = k_i +q
\end{array}
\labell{bother3}
\eeq
where $q=\frac{2}{3} \tk + \frac{1}{3} \sum_{i=2}^5 k_i$.  A
particular case appears for $p=5$ with $\gamma=\theta$,
$k_2=k_3=k_4=k_5$, $\tk=-2k_2$, $k_6=-k_1$, $k_1=\sqrt{3}$, $H=G=1$.
This solution corresponds to a six-dimensional self-dual string and
provides a nontrivial example of a non-dilatonic SD-brane.

Finally one can consider the case of SD$p$/SD($p$--6)-branes.
Actually this category only naturally includes the SD6/SD0-brane.
From experience with the D6/D0-brane bound states, we expect that the
general solution for these four-dimensional dyons will be more
complicated than the previous bound states \cite{finn} and will
require a more elaborate generating solution than given in
eq.~\reef{boso} \cite{finn2}.  However, following \cite{Khuri}, one
can realize that certain special cases should correspond to solutions
of the Einstein-Maxwell equations in four spacetime
dimensions. Precisely, such an S0-brane in four dimensions was
presented in~\cite{space} and so that solution can be embedded in type
IIa string theory. Upon lifting this solution to eleven dimensions, it
becomes a purely gravitational solution of low energy M-theory
equations. A slightly generalized solution has the eleven-dimensional
metric:
\beqa
ds^2 &=& \gamma^2 \left(-dt^2+t^2(d\chi^2+\sinh^2\chi\,d\lambda^2)
       \right) + \left(\frac{\alpha\beta}{\gamma}\right)^2 dx^2 +
\nonumber\\
&&\qquad\qquad +\left( dz +\frac{2\sqrt{2}\,\om\sin\theta }{\gamma t} dx
 + 2\sqrt{2}\,\om \sin\theta\cosh\chi\,d\lambda\right)^2 + dx_{[6]}^2
\eeqa
where as before $\alpha=1+\om/t$ and $\beta=1-\om/t$. We have also
introduced $\gamma=1+2\cos\theta\,\om/t+\om^2/t^2$.  When
$\sin\theta=0$, $\gamma=\alpha^2$ and the solution reduces to the form
given in eq.~\reef{boso}. On the other hand, $\cos\theta=0$ yields the
M-theory lift of the solution presented in \cite{space}. This
concludes our survey of multiply charged S-branes.

%--------------------------------------------------------------------+
\section{The IIb SD(--1)-brane}

As discussed above, our construction of SD$p$-brane solutions applied
in a straightforward way for $0\le p\le6$. The cases $p$=7 and 8 were
considered in the discussion section. One might also wonder if there
is such an object as an SD(--1)-brane. Recall that the usual
D(--1)-brane is an instanton carrying a $C^{(0)}$ `charge' in ten
Euclidean dimensions. This object is associated with a quantum
tunneling process in the type IIb theory. In contrast, an
SD(--1)-brane would be associated with a real-time decay process in
ten-dimensional Minkowski space. In common with its instantonic
counterpart, however, it should also carry a RR flux of
$F^{(9)}=*dC^{(0)}$. While the standard construction presented in
section 2 does not apply for $p$=--1, one can construct a candidate
type IIb supergravity solution as follows:

Begin with the eleven-dimensional solution in eq.~\reef{boso} with
$p=-1$, \ie no $x^i$'s. Now dimensionally reducing on $z$ as usual
yields:
\beqa
ds^2 &=& \ba \alpha^\half (-dt^2+t^2 dH_9^2)
\nonumber\\
e^{2\Phi} &=& \ba^{3}
\labell{stepone}
\eeqa
where we already have imposed the constraints \reef{solve}.  This type
IIa solution only involves NS fields and so is equally valid as a low
energy solution of type IIb supergravity.  Now regarding
eq.~\reef{stepone} as a type IIb solution, we apply the SL(2,R)
transformation \cite{john}
\beq
\left(\begin{array}{cc} \sin\gamma & -\cos\gamma \\
\cos\gamma & \sin\gamma\end{array}\right)
\labell{transf}
\eeq
which is chosen so that the asymptotic values of $C^{(0)}$ and
$e^{2\Phi}$ remain $0$ and $1$, respectively.  The resulting solution,
which is our candidate for the SD(--1)-brane, takes the form
\beqa
ds^2 &=& F^{\half} \alpha \beta^{-\half} (-dt^2+t^2 dH_9^2)
\nonumber\\
e^{2\Phi} &=& F^2 \ba^{-3}
\labell{lastsol}\\
C^{(0)} &=& \sin\gamma\cos\gamma \frac{C}{F}
\nonumber
\eeqa
where we have introduced the functions
\beq
C(t) = 1-\ba^3\ ,
\qquad\qquad
F(t) = \cos^2\gamma + \sin^2\gamma \ba^3\ .
\labell{lastt}
\eeq
%

%--------------------------------------------------------------------+

%\errstop %useful error on Rob's laptop

\end{document}